\documentclass[twocolumn,showpacs,preprintnumbers,amsmath,amssymb]{revtex4}
\usepackage{graphicx}
\usepackage{dcolumn}
\usepackage{bm}
\usepackage{color}
\usepackage[all]{xy}
\usepackage{amsmath}

\begin{document}
\input epsf

\title{Packing defects and the width of biopolymer bundles}
\author{Nir S. Gov}
\address{Department of Chemical Physics,
The Weizmann Institute of Science,\\
P.O.B. 26, Rehovot, Israel 76100}

\begin{abstract}
The formation of bundles composed of actin filaments and
cross-linking proteins is an essential process in the maintenance of
the cells' cytoskeleton. It has also been recreated by in-vitro
experiments, where actin networks are routinely produced to mimic
and study the cellular structures. It has long been observed that
these bundles seem to have a well defined width distribution, which
has not been adequately described theoretically. We propose here
that packing defects of the filaments, quenched and random,
contribute an effective repulsion that counters the cross-linking
adhesion energy and leads to a well defined bundle width. This is a
two-dimensional strain-field version of the classic Rayleigh
instability of charged droplets.
\end{abstract}

\maketitle

\input epsf

Filamentous biopolymers, such as F-actin, have the ability to
cross-link into a variety of bundles and networks, forming the
cytoskeleton. The distribution of the radii of cross-linked actin
bundles is a basic characteristic that determines the mechanical
properties of the cytoskeleton. The thickness of bundles similarly
determines the mechanical properties of artificial networks that
form in-vitro \cite{invitronetwork,bauschdiamater}. Recent in-vitro
experiments \cite{bernheim} indicate a broad distribution of radii,
with a distinct peak, while simple equilibrium theory would predict
a global phase separation and formation of a single bundle of
infinite width and length \cite{itamar} (in an infinite system).
Several possibilities have been proposed to explain the observed
distribution; When the bundles form due to multi-valent ions,
electrostatic interactions are unbalanced and can lead to a
selection of an equilibrium finite radius \cite{pincushenle}. A
recent study attributes the finite size of the actin bundles to the
inherent chirality of these filaments \cite{bruinsma2007}, and
indeed this may be the dominant effect when the filaments are linked
by small multi-valent ions \cite{helicalpacking,safinya2007}. We
will deal here with neutral systems that are strongly chemically
cross-linked by larger linker proteins
\cite{bernheim,wong,bauschdiamater}. In these cases the chirality of
the individual filaments may play a lesser role, and we will treat
the filaments as achiral. Our model proposes therefore a
complimentary mechanism for finite bundle widths, which applies to
achiral bundles. We present here a model where quenched disorder, in
the form of twist-defects (Fig.1), leads to the selection of a
quasi-equilibrium finite radius. We call this a quasi-equilibrium
since the defects in the bundle can be annealed away in principle,
but this is highly unlikely for long filaments and strong
cross-linking, and the system is therefore dynamically arrested in a
meta-stable configuration. There is recent experimental evidence
that in-vitro actin bundles do indeed have such twist defects
\cite{wong}. We wish to consider the implication of such defects for
the width distribution of in-vitro actin bundles. We will consider
these knots to be in a form of quenched, i.e. static, and random
disorder.

Consider a bundle of cross-linked filaments, as shown in Fig.1. The
filaments are assumed to be tightly cross-linked to all their
neighbors, so that they form a solid structure. If all the filaments
are perfectly aligned then they form a hexagonal close-packed
crystal. But in the process of aggregation twist-defects can form,
where two filaments attach to the growing bundle in a different
relative orientation along their lengths (Fig.1a). If these two
filaments stay on the surface of the bundle and have a single such
twist along their length, then this twist can slide to the ends of
the filaments and relax away. When the filaments are long enough it
is likely that they each form such twist-defects with other
neighboring filaments in other places along their lengths, and the
whole bundle is in fact knotted (Fig.1a). The knots can not be
easily relaxed now by sliding them to the end of the filaments, as
they are all entangled. The meandering of the filaments in the
bundle due to the twist-defects may be described as a random walk of
the filament, where the average number of defects along its length
is: $N_{def}\sim(L/L_p)P_{def}$, where $L$ is the overall length of
the filament, and each twist-defect extends over a length $L_p$
which is of the order of the persistence length of actin in the
bundle ($L_p\sim16\mu$m for free actin filaments
\cite{actinpersist}), and $P_{def}$ is the probability of exciting a
defect.

When actin bundles attract each other, through the strong adhesion
of cross-linking proteins, they begin to aggregate laterally. We
will consider here the case of strong cross-linking and long
filaments, so that the interaction term dominates over the entropy
term in the free energy, which we will neglect from now on
\cite{pincushenle}. The adhesion energy gain is driving the
aggregation, and since the filaments on the surface have less
adhesion energy (less neighbors), there is an energetic drive to
increase the bundle thickness. The adhesion energy per unit length
can be simply written as
\begin{equation}
E_{bind}\approx -\pi (R/a) \varepsilon_{b}((R/a)-2) \label{ebind}
\end{equation}
where $\varepsilon_{b}$ is the adhesion energy per unit length,
between the filaments due to the presence of the cross-linking
proteins. The first term represents the bulk adhesion energy
(negative) and the second term represents the surface energy
(positive). This energy functional gives a critical radius of $R=a$
($a$ is the radius of the individual filaments and surrounding
cross-linkers, which is typically \cite{wong,bernheim} $\sim10$nm),
beyond which the bundle grows to infinite width. The observed widths
of actin bundles seem to be rather well-defined \cite{wong}, and
despite the mobility of the filaments the bundles only grow by
longitudinal aggregation and do not continue to thicken through
lateral aggregation. This observation is therefore at odds with the
equilibrium theory described by Eq.(\ref{ebind}), and is our
motivation for looking at the effect of packing defects.

We now treat the case of twist-defects inside the bundle. Each
defect involves an increase in the local energy due to several terms
(Fig.1): (i) Loss of adhesion energy due to broken cross-linking
bonds at the site of the twist, (ii) the elastic energy of the
twisted actin filaments, (iii) the elastic energy of the deformed
hexagonal lattice of filaments around the defect in the twist plane
(Fig.1). Out of these energies the first and third can be different
if the defect is inside the bulk of the bundle or close to the
surface. Furthermore, it is clear that these two terms are in fact
\emph{smaller} at the surface, so the energetic cost of a filament
is higher inside the bulk of the bundle. A uniform distribution of
static defects will therefore cost a higher energy the thicker is
the bundle. We can already see how the defects may compete with the
adhesion energy and drive the bundles to a finite preferred width.

Let us treat the bundle as being approximately uniform along its
length, so that we can treat the defects in a two-dimensional
circular cross-section. The strain field in the surrounding bundle
due to a localized defect depends on the order of the defect, i.e. a
monopole or a higher multipole, and is maximal in the maximal
twist-plane (Fig.1). A twist of two filaments increases the local
volume of the filament packing (Fig.1), so has a monopole component
(dilation). At the same time the twist involves also a quadropole
component, as shown in Fig.1c. The single isolated monopole defect
gives rise to a strain field that decays as $u\sim1/r$ away from the
defect, and the resulting strain energy has a logarithmic divergence
with the radius of the bundle. The contribution from such a single
defect is therefore negligible compared to the $R^2$ term in
Eq.(\ref{ebind}), and the equilibrium stays at an infinite width.

We are therefore led to consider a uniform distribution of many
defects. Before dealing with the complications of the long-range
strain field of the monopole, let us treat the case of a highly
localized strain-field, such as for the quadropole. In this case we
can treat the strain energy in the bundle around the defect as part
of the defect core energy $E_c$, that includes the broken
cross-linker bonds and twist of the actin filaments.

A uniform distribution of defects, of density $\rho=1/L^2$, will
simply add $E_{defect}=\rho \pi R^2 E_{c}$ to the energy of
Eq.(\ref{ebind}), without any qualitative change, i.e. the system
still has minimum energy for an infinitely wide bundle. Let us
assume that close to the rim of the bundle defects are not strongly
trapped, and can relax. Such a process may occur due to higher
mobility of the filaments on the bundle surface, which leads to
effective "surface-melting" of several layers of filaments, thereby
annealing any defects. The thickness of this annealed layer is
denoted by $\lambda$ (Fig.2a). The energy of the defects (per unit
length) in this system is therefore given by
\begin{equation}
E_{defect}\approx \rho \pi (R-\lambda)^2 E_{c} \label{edefectempty}
\end{equation}

The total energy per unit length is now
$E_{total}=E_{bind}+E_{defect}$
(Eqs.\ref{ebind},\ref{edefectempty}). We find that there is a
critical value of the defect energy $E_c^*=\varepsilon_{b} (L/a)^2$,
above which the system has an equilibrium configuration of finite
width, given by
\begin{equation}
R_{0,shell}=(E_{c}\lambda-aE_c^*)/(E_{c}-E_{c}^*) \label{r0shell}
\end{equation}
Note that as the defect core energy increases above its critical
value, the bundle radius shrinks, for a fixed density. There is also
a critical value of the width of the annealed layer, given by:
$\lambda_{c}=aE_c^*/E_{c}$, at which the equilibrium bundle size
shrinks to zero. Above $E_c^*$ the system is dominated by the energy
of the defects, while below it the binding energy dominates. In the
limit of $E_{c}\gg\varepsilon_{b},E_c^*$, we find that the
equilibrium radius approaches the thickness of the defect-free layer
$R_{0,shell}\rightarrow\lambda$, while in the limit
$E_{c}\rightarrow E_c^*$ the equilibrium radius diverges.

When the width of the defect-free outer shell is given by the
typical inter-defect distance, i.e. $\lambda=L$, we get an
equilibrium configuration of finite width, given by
(Eq.\ref{r0shell}):
$R_{0,shell}=L(E_{c}-\varepsilon_{b}L/a)/(E_{c}-E_{c}^*)$, and a
critical inter-defect separation of:
$L_{c}=a(E_{c}/\varepsilon_{b})$.

We now return to the problem of a uniform distribution of
strain-monopoles. These can be treated as a two-dimensional gas of
charges, all of the same sign, interacting via logarithmic repulsion
\cite{thouless} (Fig.2b, bottom). Note that since the twist-defects
cause a local dilation, they all have the sam sign and repel each
other. The problem therefore resembles a two-dimensional version of
the famous Rayleigh-instability of charged droplets \cite{rayleigh}.
In our case the twist-defects behave as elastic charges and their
mutual repulsion breaks the infinitely thick bundle into bundles of
finite radius.

We can estimate the overall strain energy (per unit length) of the
system of uniform monopoles by calculating the interaction of a
central charge with a uniform distribution of surrounding charges,
assuming that its nearest neighbor is a distance $L$ away (Fig.2b)
\begin{eqnarray}
E_{mon}&\approx& k \ln{(L/a)}+2 k \rho \int_L^R \ln{(r/a)}d^2r
\label{emonopole}\\
=k&\ln{(L/a)}&+ 2 k
\pi\left(1-\frac{R^2}{L^2}-\ln{(L/a)}+\frac{R^2}{L^2}\ln{(R/a)}\right)
\nonumber
\end{eqnarray}
where $k$ is the elastic modulus of the bundle which is of order
$\varepsilon_{b}$. The first term in Eq.(\ref{emonopole}) is the
energy to create the defect at a distance $L$ from its nearest
neighbor, and the second is the interaction energy with a uniform
distribution of surrounding defects. We find that the dominant term
in $R$ is of order $(R/L)^2\ln{(R/a)}$. Since we do not attempt an
exact solution of the complex strain-field inside a cylindrical
bundle with defects, we will continue with a scaling analysis using
this term, i.e. we will approximate the strain-field energy per unit
length due to the defects as
\begin{equation}
E_{defects}\approx 2\pi k' \frac{R^2}{L^2}\ln{(R/a)} \label{defects}
\end{equation}
where $k'$ combines the effective stiffness of the bundle and
various geometric factors. This energy dominates at large $R$ over
the binding energy (Eq.\ref{ebind}), due to the $\ln{(R/a)}$ factor.
We therefore have a situation now where the bundles always have a
finite equilibrium radius.

The total energy per unit length is now
$E_{total}=E_{bind}+E_{defects}$ (Eqs.\ref{ebind},\ref{defects}),
and has a global minimum at the equilibrium radius, given by
\begin{equation}
R_{0,mono}/a= \exp{((1/2a^2\rho \alpha)-1/2)} \label{r0mono}
\end{equation}
where $\alpha\equiv k'/\varepsilon_{b}$. We find that as the density
of defect increases the radius decreases, while it also decreases
with the stiffness of the bundle $\alpha$. Note that the equilibrium
radius has an exponential dependence on the density of defects, and
is therefore predicted to be very sensitive to this parameter. The
calculation we have given above are all appropriate for a system
with an infinite reservoir of actin filaments and cross-linking
proteins. In a closed system with a finite number of available
filaments and cross-linkers we have to conserve the overall area of
actin bundles, which should not change the overall behavior we
described.

If the defects are interacting strongly with each-other through
their \emph{quadropole} component of the strain, then they can
arrange in string-like aggregates (Fig.1e), in the maximal
twist-plane. These aggregates can form if the density of defects is
high and they are mobile enough to move inside the bundle and
aggregate. Alternatively such aggregates of defects may form as the
filaments join the bundle at the surface, since a defect is more
likely to form close to an existing defect; at the right orientation
it has a lower strain energy (Fig.1d). Such aggregations of defects
will modify the strain-field and the contribution of the defect
energy, but is beyond the scope of this paper.

Let us now compare our calculation with recent in-vitro measurements
of the distribution of actin bundle radii \cite{bernheim}. The size
distribution of bundles can be approximated to follow from a simple
Boltzman distribution of the bundle energy:
$P(R)\propto\exp{(-(E_{total}(R)-E_0)/k_{B}T}$. Since the bundle
energy has a quadratic minimum at $R_0$
(Eqs.\ref{r0shell},\ref{r0mono}), this gives a Gaussian distribution
of radii centered around the equilibrium radius. The distribution of
the number of actin filaments inside the bundles $N$ is simply the
distribution of bundle areas, $N\propto (R/a)^2$, which we we
predict to have roughly an exponential behavior at large $N$ since
both Eqs.(\ref{edefectempty},\ref{defects}) are quadratic with $R$
(for the monopoles there is a slight logarithmic correction). This
is in good agreement with the observed distribution (Fig.3).

The width of the distribution can be estimated from the above
calculation, using
\begin{eqnarray}
\langle \Delta R^2\rangle&=&\frac{k_{B}T}{\partial^2
E_{total}/\partial R^2}|_{R_0} \nonumber \\
&=&\frac{k_{B}T}{2 \rho(
E_c-E_c^*)},\quad\frac{k_{B}T}{4k'\rho}\label{noise}
\end{eqnarray}
where the first is for the empty shell case
(Eqs.\ref{edefectempty},\ref{r0shell}), and the second is for the
uniform field of monopoles (Eqs.\ref{defects},\ref{r0mono}). We find
that as the density of defects increases the width of the bundle
sizes distribution decreases, and the distribution becomes tighter
around $R_0$.

In Fig.3 we plot the calculated distribution compared to the
experimental measurement \cite{bernheim}, at a ratio of 1:5 Fascin
cross-linkers to actin in the solution. In these fits we have to
choose the unknown binding energy, and then fit the value of the
parameters so that we get the position of the peak of the
distribution and its width (Eq.\ref{noise}). There is some freedom
in the choice of these parameters; for the empty-shell model we
choose the values; $\varepsilon_{b}=10k_{B}T/a$ which is of the
order of the known adhesion energy per typical actin cross-linker
\cite{itamar}, and $a^2\rho=0.04$ which corresponds to a rather
dense array of defects and is quite arbitrarily. By fitting to the
experimental distribution, we then fix the values of
$\lambda/a=1.007$ and $E_c=250k_{B}T/a$.

For the uniform-monopoles the width of the observed distribution
fixes the value of $a^2\rho k'$, and we take $a^2\rho=0.1$ and
$k'=0.03k_{B}T/a$, which then fixes the value of
$\varepsilon_{b}=0.05k_{B}T/a$, by fitting to the peak location. We
find that for a good fit we need a very small value for the binding
energy per unit length. The reason for this may arise from the fact
that our two-dimensional monopoles represent twist-defects which
extends over a length of the order of the persistence length of
actin $L_p$ \cite{actinpersist}. This may be much longer than the
effective thickness of the maximal twist-plane $L_{mono}$ (Fig.1).
If $L_{mono}\sim a$ then all the energy terms are re-scaled by a
factor of $L_{mono}/L_p\sim10^{-2}-10^{-3}$.

It has been further observed that the average thickness of the
bundles increases with the concentration of cross-linking proteins
\cite{bauschdiamater,bernheim}. Within our model the radius of the
peak of the distribution, at $R_{0,shell}$ or $R_{0,mono}$, depends
only on the density of defects $\rho$ if both the binding energy
$\varepsilon_{b}$ and the defect energies $E_c,k'$ depend in the
same way on the concentration of cross-linking proteins (see
Eqs.\ref{edefectempty},\ref{r0shell}). We can therefore propose that
as the concentration of cross-linking proteins increases the density
of defects decrease, leading to an increase in the average radius of
the bundles; both the peak position and the width of the
distribution (Eq.\ref{noise}) increase with decreasing $\rho$. The
more weakly bound bundles appear therefore to aggregate in a
"messier" fashion, resulting in more twist-defects. The probability
of exciting a defect $P_{def}\sim\exp{(-\Delta E/k_{B}T)}$ per unit
length of filament, is higher when the energy barrier to create a
defect at the bundle surface $\Delta E$, decreases with the decrease
in the concentration of cross linkers. This prediction may be
checked in future experiments.

In a living cell the actin filaments form bundles in a variety of
forms. Some bundles are inside the bulk of the cell cytoplasm, and
their formation may therefore resemble that of in-vitro bundles, and
consequently their width may be limited by the appearance of
defects. Another form of actin bundles appear in the core of
stereocilia, and seem to be packed in a perfectly regular, hexagonal
lattice \cite{bechara2004}. These bundles do not form by the lateral
aggregation of preexisting long filaments as in the bulk in-vitro
experiments; all the filaments in such a bundle polymerize together
and elongate in synchrony from the stereocilia tip. The
polymerization of the filaments at their ends is promoted by
tip-complex proteins. If the relative locations of the nucleation
sites within this tip complex are maintained in a solid-like static
structure, then the filaments which elongate from these nucleation
sites will form a perfectly packed bundle. The width of these
bundles is therefore not controlled by the random process of defect
formation, but is very tightly controlled by the cell
\cite{bechara2004}. Indeed such bundles contain thousands of actin
filaments, much thicker than the in-vitro bundles (Fig.3).

We conclude that the aggregation of long filaments by strong
cross-linkers is likely to produce packing defects. Such quenched
and random defects can lead to an effective pressure that limits the
growth of the width of bundles, and may explain the observed peaked
distribution with an exponential tail. It is a two-dimensional
elastic version of the Rayleigh-instability of charged droplets
\cite{rayleigh}. The rudimentary treatment given here should be
improved in the future by a more rigorous treatment of the full
three-dimensional problem of the complex strain-fields of a
population of twist-defects inside a bundle that is free to twist
and bend as a whole.

\begin{acknowledgments}
I thank the Alvin and Gertrude Levine Career Development Chair, for
their support. This research was supported by the Israel Science
Foundation (grant No. 337/05).  This research is made possible in
part by the historic generosity of the Harold Perlman Family.
\end{acknowledgments}

\begin{figure}
\centerline{\ \epsfysize 10cm \epsfbox{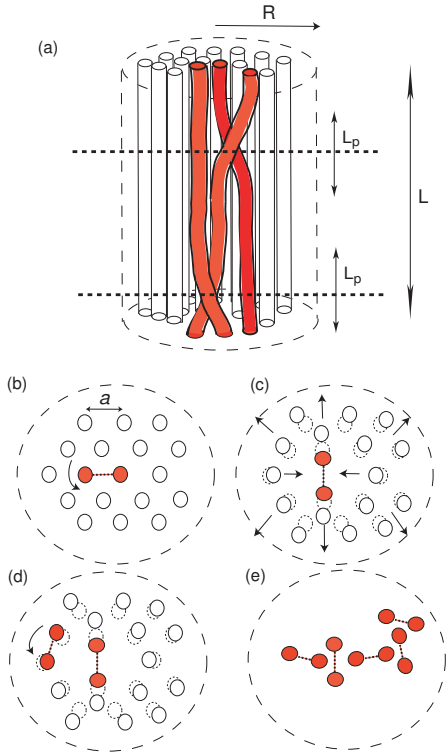}}
\caption{(a) Schematic picture of a single twist-defect (red) inside
a bundle of radius $R$ and overall length $L$. The twist extends
over a length of the order of the persistence length of actin
$L_{p}$. Several twist-defects along the length of the bundle cause
the filaments to form a knot. Far below or above the maximal twist
plane (horizontal dashed line), the filaments are arranged in a
two-dimensional hexagonal array (b). (c) In the maximal twist-plane
the defect (red) creates a strain field around it (arrows), and
pushes the surrounding filaments from their perfectly hexagonal
array (dashed circles). (d) The quadropole strain field around a
defect lowers the energy to create a near-by defect with orthogonal
orientation (arrow), and can result in a string-like aggregate of
defects (e).}
\end{figure}

\begin{figure}
\centerline{\ \epsfysize 4cm \epsfbox{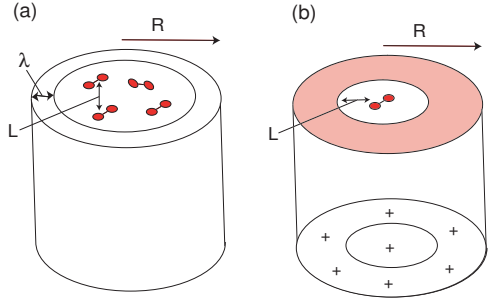}}
\caption{(a) Schematic picture of the empty-shell model; uniform
population of defects (red pairs) at areal density $\rho=1/L^2$ and
core energy per unit length $E_c$, with an empty outer shell of
thickness $\lambda$. (b) Schematic calculation of the uniform field
of monopoles; a single monopole (red pair in the center) interacts
with a uniform distribution of neighboring monopoles of density
$\rho$ (red ring), beyond a minimal separation radius $L$. This
system is the elastic equivalent to a two-dimensional electrostatic
repulsion problem (bottom).}
\end{figure}

\begin{figure}
\centerline{\ \epsfysize 6cm \epsfbox{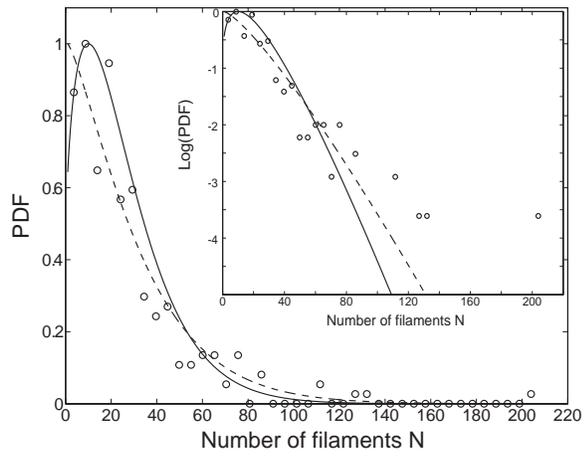}}
\caption{Experimental distribution of the radius of actin filaments
inside bundles (circles) \cite{bernheim}, for Fascin/Actin ratio of
1:5. The calculated distribution is given by the solid line for the
empty-shell model (Eq.\ref{edefectempty}) and dashed line for the
uniform monopoles (Eq.\ref{defects}). The inset gives a semi-log
plot of the distribution.}
\end{figure}

\end{document}